\documentclass[twocolumn,showpacs,preprintnumbers,a4,prb]{revtex4}
\usepackage{graphicx}
\usepackage{dcolumn}
\usepackage{amsmath}
\usepackage{amssymb}
\usepackage{bm}

\begin{document}

\title{Stray fields based magnetoresistance mechanism in Ni$_{80}$Fe$_{20}$-Nb-Ni$_{80}$Fe$_{20}$ trilayers}

\author{D. Stamopoulos,\cite{cor} E. Manios, and M. Pissas}

\affiliation{Institute of Materials Science, NCSR "Demokritos",
153-10, Aghia Paraskevi, Athens, Greece.}
\date{\today}

\begin{abstract}
In this work we report on the transport and magnetic properties of
hybrid trilayers (TLs) and bilayers (BLs) that consist of low spin
polarized Ni$_{80}$Fe$_{20}$ exhibiting in-plane but no uniaxial
anisotropy and low-T$_c$ Nb. We reveal a magnetoresistance effect
of magnitude identical to the one's that were reported in [V.
Pe\~{n}a et al., Phys. Rev. Lett. {\bf 94}, 57002 (2005)] for TLs
consisting of highly spin polarized La$_{0.7}$Ca$_{0.3}$MnO$_3$
and high-T$_c$ YBa$_2$Cu$_3$O$_7$. The presented effect is
pronounced when compared to the one reported in [A.Yu. Rusanov et
al., Phys. Rev. B {\bf 73}, 060505(R) (2006)] for
Ni$_{80}$Fe$_{20}$-Nb-Ni$_{80}$Fe$_{20}$ TLs of strong in-plane
uniaxial anisotropy. In our TLs the magnetoresistance exhibits an
increase of two orders of magnitude when the superconducting state
is reached: from the conventional normal-state values $\Delta
R/R_{nor}\times100\%=0.6\%$ it goes up to $\Delta
R/R_{nor}\times100\%=45\%$ ($\Delta R/R_{min}\times100\%=1000\%$)
for temperatures below T$_c^{SC}$. In contrast, in the BLs the
effect is only minor since from $\Delta R/R_{nor}\times100\%=3\%$
in the normal state increases only to $\Delta
R/R_{nor}\times100\%=8\%$ ($\Delta R/R_{min}\times100\%=70\%$) for
temperatures below T$_c^{SC}$. Magnetization data of both the {\it
longitudinal} and {\it transverse} magnetic components are
presented. Most importantly, in this work we present data not only
for the normal state of Nb but also in its superconducting state.
Strikingly, these data show that {\it below its T$_c^{SC}$ the Nb
interlayer under the influence of the outer Ni$_{80}$Fe$_{20}$
layers attains a magnetization component transverse to the
external field.} By comparing the transport and magnetization data
we propose a new candidate mechanism that could motivate the
pronounced magnetoresistance effect observed in the TLs. Adequate
magnetostatic coupling of the outer Ni$_{80}$Fe$_{20}$ layers is
motivated by stray fields that emerge naturally in their whole
surface due to the multidomain magnetic structure that they attain
near coercivity. Consequently, the stray fields penetrate the Nb
interlayer and suppress its superconducting properties by
primarily (secondarily) exceeding its lower (upper) critical
field. Atomic force microscopy is employed in order to examine the
possibility that such magnetostatic coupling could be promoted by
interface roughness. Referring to the BLs, although out-of-plane
rotation of the single Ni$_{80}$Fe$_{20}$ layer is still observed,
in these structures magnetostatic coupling doesn't occur due to
the absence of a second Ni$_{80}$Fe$_{20}$ one so that the
observed magnetoresistance peaks are only modest.

\end{abstract}


\maketitle

\pagebreak

\section{Introduction}

Recently, superconducting-ferromagnetic (SC-FM) hybrids have
attracted much interest due to the general expectation that both
interesting features and subsequent application devices could
emerge owing to the deliberate mixing of these two long-range
phenomena.\cite{Gu02,Moraru06,Pena05,Rusanov06,StamopoulosPRB04,StamopoulosPRB05,StamopoulosPRB06,StamopoulosSST06,StamopoulosPRB06new,Volkov03,Eschrig05L,Allsworth02,Maleki06}
A very representative elemental apparatus is the superconducting
spin valve that was theoretically proposed in
Refs.\onlinecite{Buzdin99,Tagirov99}. It is based on a FM-SC-FM
trilayer (TL) where, as it was proposed \cite{Buzdin99,Tagirov99}
the nucleation of superconductivity can be controlled by the {\it
relative} magnetization orientation of the outer FM layers. J.Y.
Gu et al. \cite{Gu02} were the first who reported on the
experimental realization of a
[Ni$_{82}$Fe$_{18}$-Cu$_{0.47}$Ni$_{0.53}$]/Nb/[Cu$_{0.47}$Ni$_{0.53}$-Ni$_{82}$Fe$_{18}$]
spin valve. I.C. Moraru et al. \cite{Moraru06} also studied a
great number of Ni-Nb-Ni TLs and observed a significantly larger
shift of the superconducting transition temperature T$_c^{SC}$
than that reported in Ref.\onlinecite{Gu02}. In these works
\cite{Gu02,Moraru06} the exchange bias was employed in order to
"pin" the magnetization of the one FM layer and it was observed
that when the magnetizations of the two FM layers were parallel
(antiparallel) the resistive transition of the SC was placed at
lower (higher) temperatures. V. Pe\~{n}a et al. \cite{Pena05} and
A.Yu. Rusanov et al. \cite{Rusanov06}, studied
La$_{0.7}$Ca$_{0.3}$MnO$_3$-YBa$_2$Cu$_3$O$_7$-La$_{0.7}$Ca$_{0.3}$MnO$_3$
and Ni$_{80}$Fe$_{20}$-Nb-Ni$_{80}$Fe$_{20}$ TLs, respectively. In
agreement to what we have also reported in our recent work
\cite{StamopoulosPRB06new} both studies \cite{Pena05,Rusanov06}
reported that the antiparallel magnetization configuration of the
FM layers suppresses superconductivity when compared to the
parallel case. Interestingly, in the work of Pe\~{n}a et al.
\cite{Pena05} a magnetoresistance effect of the order $1000\%$ was
observed in the TLs that it was related to the occurrence of spin
imbalance \cite{Takahashi99} in the SC ultimately motivated by the
high spin polarization (almost $100\%$) of
La$_{0.7}$Ca$_{0.3}$MnO$_3$.

An alternative underlying mechanism is proposed in the present
work for the interpretation of a similar effect having identical
magnitude for the case of TLs constructed by two completely
different FM and SC ingredients: low spin polarized
Ni$_{80}$Fe$_{20}$ and low-T$_c$ Nb. We note that the
Ni$_{80}$Fe$_{20}$ layers employed in this work were deposited at
{\it room temperature} with {\it no} magnetic field applied during
the deposition (see the preparation details below). They exhibit
strong in-plane but no uniaxial anisotropy. We observed that for
T$<$T$_c^{SC}$ the Ni$_{80}$Fe$_{20}$-Nb-Ni$_{80}$Fe$_{20}$ TLs
exhibit an extreme zero-field increase of the measured MR which
amounts to $\Delta R/R_{nor}\times100\%=45\%$ ($\Delta
R/R_{min}\times100\%=1000\%$). Beside the TLs more simple
Nb-Ni$_{80}$Fe$_{20}$ bilayers (BLs) have also been studied. In
these BLs the effect under discussion is minimum since it amounts
to only $\Delta R/R_{nor}\times100\%=8\%$ ($\Delta
R/R_{min}\times100\%=70\%$). In the normal state of the SC
interlayer both TLs and BLs present conventional magnetoresistance
which by no means exceeds $0.6\%$ and $3\%$, respectively.

Magnetization measurements of both the {\it longitudinal} and {\it
transverse} magnetic components (in respect to the external
magnetic field) revealed that the TL exhibits almost completely
reversible longitudinal magnetization with nearly zero remanence,
while its respective transverse component attains significant
values near zero field indicating a magnetostatic coupling of the
outer Ni$_{80}$Fe$_{20}$ layers. {\it Most importantly, these
results clearly show that below its T$_c^{SC}$ the Nb interlayer
attains a magnetization component transverse to the external
magnetic field under the influence of the outer Ni$_{80}$Fe$_{20}$
layers.} These magnetic characteristics indicate that in our
samples the magnetoresistance effect is motivated by a transverse
magnetization component related to the stray fields that near
coercivity emerge naturally in the whole surface of the outer
Ni$_{80}$Fe$_{20}$ layers due to the multidomain magnetic
structure that they attain. As these stray fields interconnect the
outer Ni$_{80}$Fe$_{20}$ layers through the Nb interlayer they
surely exceed its lower or even {\it locally} amount to its upper
critical field. Accordingly, current induced vortex motion (if
only the SC's lower critical field is exceeded) or {\it localized}
normal state areas (if even the SC's upper critical field is {\it
locally} overstepped) could motivate the observed dissipation,
respectively. Since it is well known that interface roughness
could promote stray-fields induced magnetostatic coupling in
relevant layered structures
\cite{Parkin98,Parkin00,Parkin00b,Egelhoff06} we also use atomic
force microscopy (AFM) in order to examine this possibility.
Referring to the BLs although out-of-plane rotation of the single
Ni$_{80}$Fe$_{20}$ layer is still observed, due to the absence of
a second one in these structures magnetostatic coupling doesn't
occur so that the observed magnetoresistance peaks are only
modest.

Finally, a comparison with current experiments is made. Referring
to the BLs we note that the smooth magnetoresistance effect that
we observe in our Nb-Ni$_{80}$Fe$_{20}$ hybrids qualitatively
resembles the one reported by V.V. Ryazanov et al. in
Ref.\onlinecite{Ryazanov03} for the case of
Nb-Cu$_{0.43}$Ni$_{0.57}$ BLs. Regarding the TLs, the main
mechanism that we propose for the interpretation of the intense
peaks that are obtained in our
Ni$_{80}$Fe$_{20}$-Nb-Ni$_{80}$Fe$_{20}$ hybrids is different from
the one proposed by V. Pe\~{n}a et al. \cite{Pena05} and by A. Yu.
Rusanov et al. \cite{Rusanov06} for
La$_{0.7}$Ca$_{0.3}$MnO$_3$-YBa$_2$Cu$_3$O$_7$-La$_{0.7}$Ca$_{0.3}$MnO$_3$
and Ni$_{80}$Fe$_{20}$-Nb-Ni$_{80}$Fe$_{20}$ ones, respectively.
Reasons responsible for the existing differences are discussed.

\section{Preparation of samples and experimental details}

The samples were sputtered on Si $[001]$ substrates under an Ar
environment ($99.999 \%$ pure). In order to eliminate the residual
oxygen that possibly existed in the chamber we performed Nb
pre-sputtering for very long times.
\cite{Stamopoulos05PRB,StamopoulosSST} During the pre-sputtering
process all the residual oxygen was absorbed by the dummy Nb since
it acts as a strong oxygen getter. This procedure has a direct
impact on the quality of the produced films.
\cite{Stamopoulos05PRB,StamopoulosSST} The Nb layers were
deposited by dc-sputtering at $46$ W and an Ar pressure of $3$
mTorr, while for the Ni$_{80}$Fe$_{20}$ (NiFe) layers
rf-sputtering was employed at $30$ W and $4$ mTorr. We should
stress that: (i) all depositions were carried out {\it at room
temperature} and (ii) {\it no} external magnetic field was applied
hereon during the deposition of the NiFe layers. However, the
samples can't be shielded from the residual magnetic fields
existing in the chamber of our magnetically-assisted-sputtering
unit. Measurements by means of a Hall sensor revealed that at the
place where the substrates are mounted the residual fields exhibit
parallel components of magnitude $10-15$ Oe at maximum. Thus, our
NiFe films exhibit in-plane anisotropy. However, they don't
exhibit detectable uniaxial anisotropy since the magnetic field
sources are placed symmetrically on the perimeter of the circular
rf-gun. The produced films have low coercive fields of order $20$
Oe. In this work we show detailed results for
NiFe(19)-Nb(50)-NiFe(38) TLs and Nb(50)-NiFe(38) BLs (in nanometer
units). Both qualitatively and quantitatively similar results were
obtained in NiFe(19)-Nb(50)-NiFe(19) TLs and Nb(50)-NiFe(19) BLs.

Our magnetoresistance measurements were performed by applying a
dc-transport current (always normal to the magnetic field) and
measuring the voltage in the standard four-point straight
configuration. In most of the measurements the applied current was
$I_{{\rm dc}}=0.5$ mA, which corresponds to an effective density
$J_{{\rm dc}}\approx 1000$ A/cm$^2$ (typical in plane dimensions
of the films are $6\times 5$ mm$^2$). The temperature control and
the application of the magnetic fields were achieved in a
superconducting quantum interference device (SQUID) (Quantum
Design). In all cases the applied field was parallel to the films.

\section{Experimental results}

\subsection{Magnetoresistance data}

\begin{figure}[tbp] \centering%
\includegraphics[angle=0,width=7.5cm]{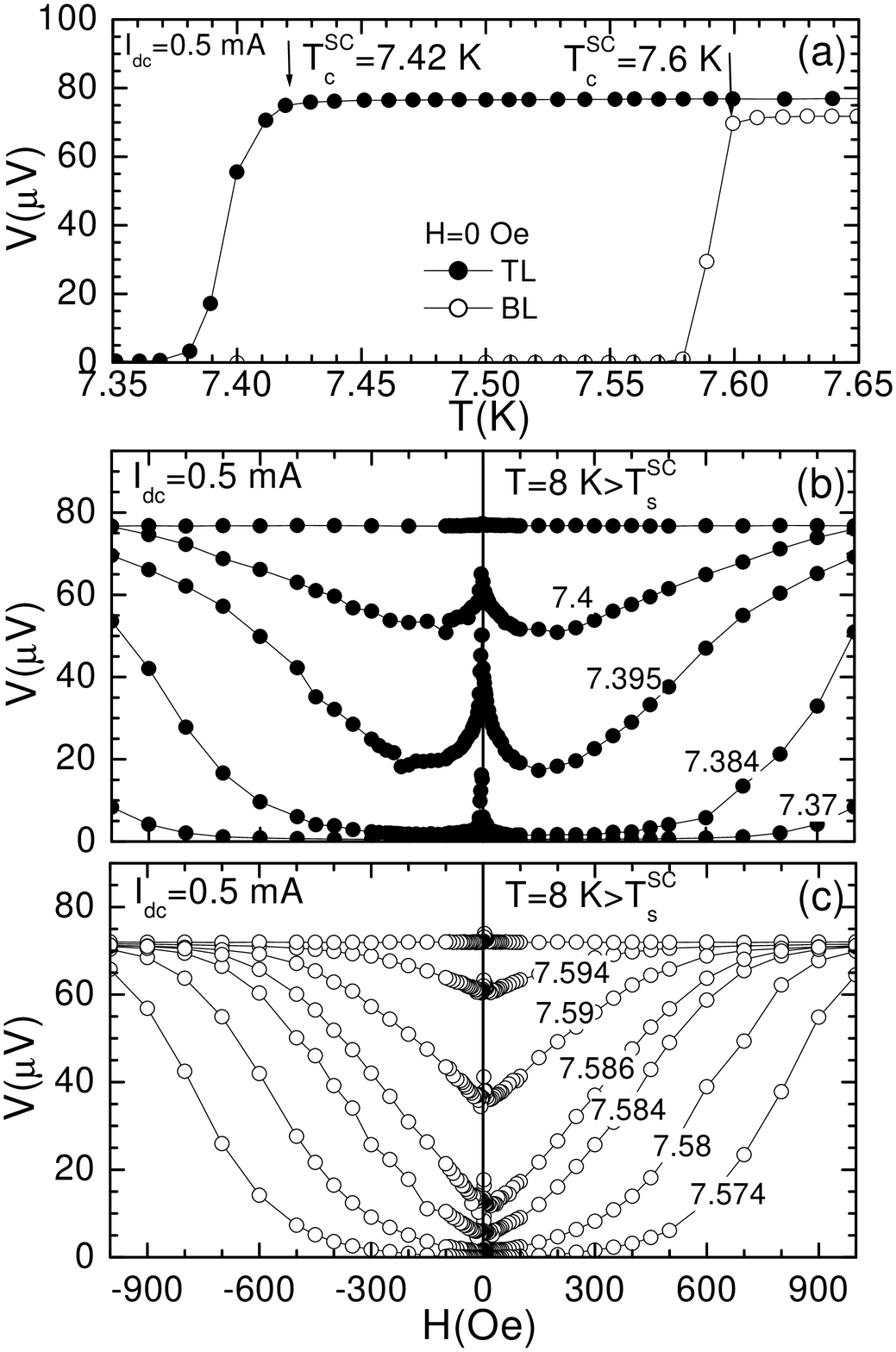}
\includegraphics[angle=0,width=7.5cm]{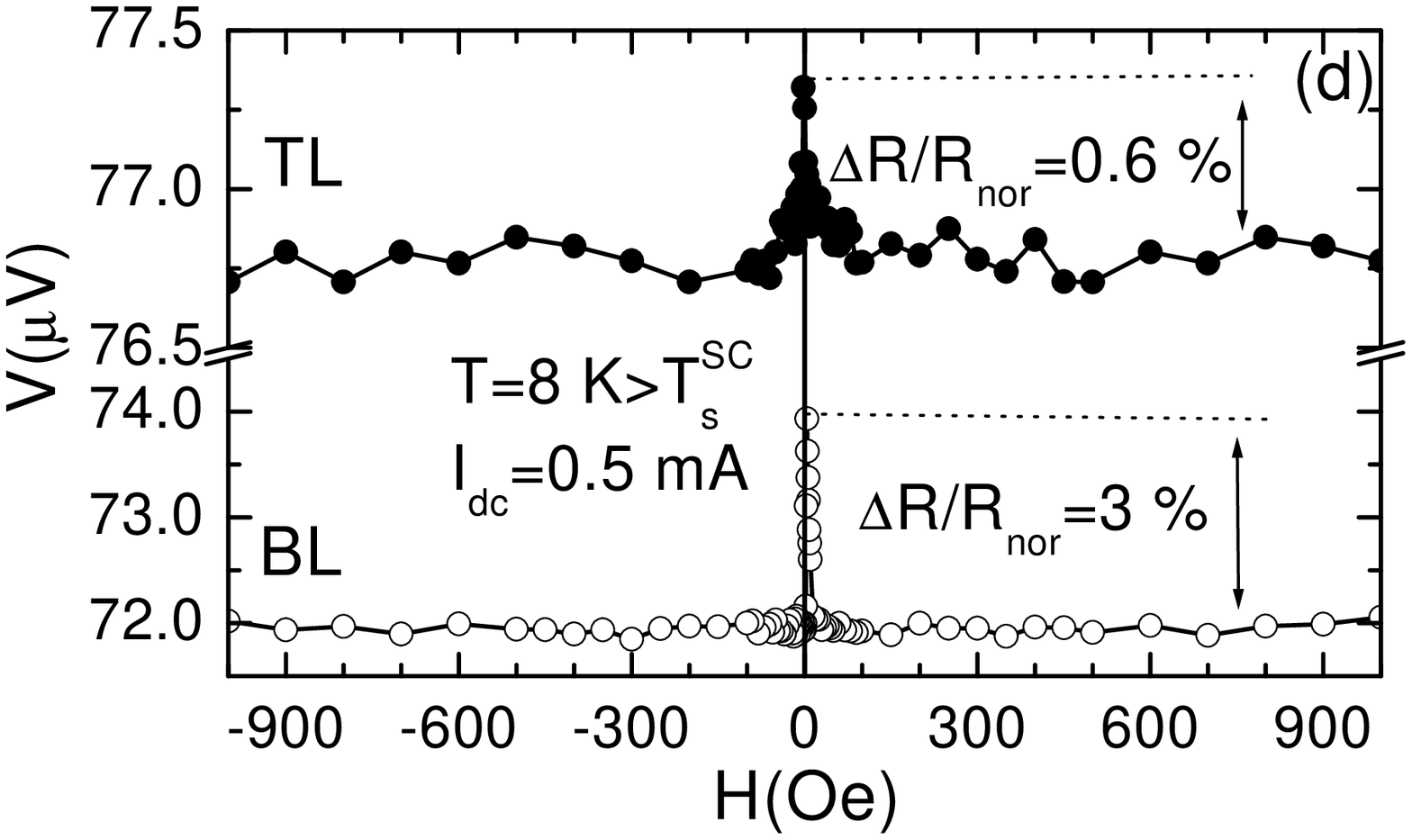}
\caption {(a) Zero-field resistive curves V(T) of a TL (solid
circles) and a BL (open circles), respectively. Detailed V(H)
curves for a TL (b) and for a BL (c) at various temperatures
across their resistive transitions presented in (a). In panel (d)
we focus on the respective curves observed in the normal state for
both the TL (solid circles) and the BL (open circles).}
\label{b1}%
\end{figure}

As may be seen in Fig. \ref{b1}(a) the zero-field critical
temperature is T$_c^{SC}=7.42$ K and T$_c^{SC}=7.6$ K for the TL
and the BL, respectively. The slightly lower critical temperature
of the TL is fairly justified since owing to the proximity effect
the two outer NiFe layers of the TL suppress stronger the
superconducting order parameter when compared to the BL. Figures
\ref{b1}(b) and \ref{b1}(c) show the main raw experimental results
of our work. Presented are detailed voltage curves V(H) as a
function of magnetic field for both the TL, panel (b) and the BL,
panel (c) that were obtained in various temperatures across their
zero-field resistive curves that are presented in panel (a). We
clearly see that the TL exhibits a completely different behavior
when compared to the BL. While as the magnetic field is lowered
from $1000$ Oe the resistance of both structures decreases, it is
the resistance of the TL that below a threshold value increases
strongly exhibiting a maximum at around zero field. The field's
regime where the resistance increase is observed in the TL extends
to no more than $300-400$ Oe. In contrast, the respective curves
of the BL present only a minor increase for $5-10$ Oe around zero
field.

\begin{figure}[tbp] \centering%
\includegraphics[angle=0,width=6.5cm]{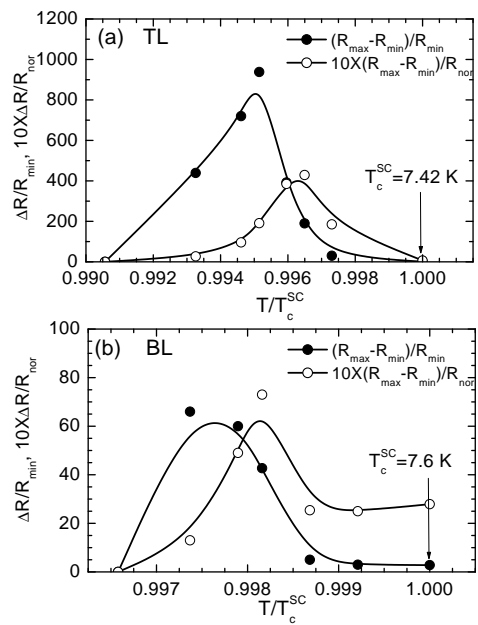}
\caption {Percentage change of the MR as calculated according to
two definitions $(R_{max}-R_{min})/R_{min}\times100\%$ and
$(R_{max}-R_{min})/R_{nor}\times100\%$ for the TL (a) and the BL
(b). In both panels the data referring to the second definition
$(R_{max}-R_{min})/R_{nor}\times100\%$ are multiplied by a factor
of $10$ for the sake of presentation.}
\label{b2}%
\end{figure}%

The magnetoresistance effect observed in the TL is strong. The
percentage resistance change $(R(0)-R(H))/R(H)\times100\%$ meets
the criteria for considered as giant magnetoresistance. Figures
\ref{b2}(a) and \ref{b2}(b) present the respective data for the TL
and the BL as a function of the reduced temperature according to
two definitions. The first one is
$(R_{max}-R_{min})/R_{min}\times100\%$ (solid circles) and takes
into account the minimum (around $150-200$ Oe) and maximum (around
$0$ Oe) values of the experimental curves. The second one, which
is modest and surely more relevant for the study of the underlying
physics, is $(R_{max}-R_{min})/R_{nor}\times100\%$ (open circles)
and takes into account the resistance that the SC exhibits in its
normal state. In Fig. \ref{b2}(a) we see that for the TL the
maximum MR is obtained halfway the resistive transition of the SC,
at T/T$_c^{SC}$=$0.995$ and amounts to almost $1000\%$ according
to the first definition. But even according to the modest
definition $(R_{max}-R_{min})/R_{nor}\times100\%$ the observed
increase amounts to $45\%$ which is well inside the range of giant
magnetoresistance (notice that for the sake of presentation in
both panels these data are multiplied by a factor of $10$). Figure
\ref{b2}(b) presents the respective data for the BL. In contrast
to the TL, for the BL the maximum magnetoresistance is obtained
near the bottom of its resistive transition, at
T/T$_c^{SC}$=$0.9973$ and doesn't exceed $70\%$ ($8\%$) according
to the first (second) definition. {\it It is obvious that the
effect observed in the TL is at least one order of magnitude
stronger than the one of the BL.} Despite the differences
discussed above we have to stress that in both structures the
effect is not observed in the normal state of the SC. More
specifically, for T$>$T$_c^{SC}$ conventional magnetoresistance is
revealed with values $0.6\%$ and $3\%$ for the TL and BL,
respectively as it is presented in Fig.\ref{b1}(d).

\subsection{Magnetization data}

\begin{figure}[tbp] \centering%
\includegraphics[angle=0,width=7cm]{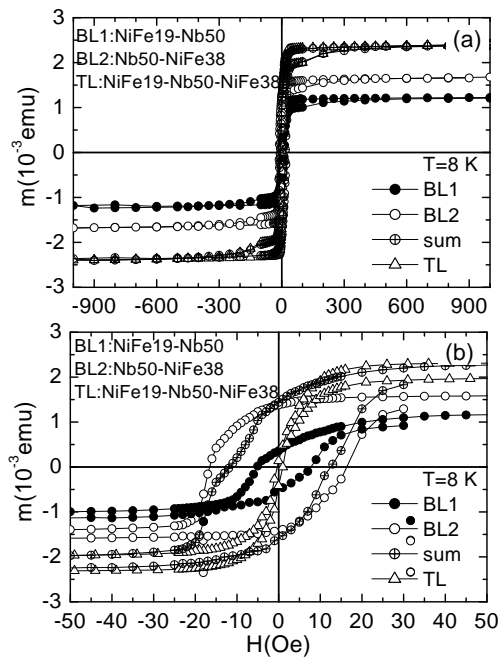}
\caption {Magnetization loop data obtained at T$=8$ K$>$T$_c^{SC}$
for the NiFe(19)-Nb(50)-NiFe(38) TL (triangles), a NiFe(19)-Nb(50)
BL1 (solid circles) and a Nb(50)-NiFe(38) BL2 (open circles).
Circles with crosses correspond to the simple addition of the m(H)
loops of the BL1 and BL2 (see text for details). All data are
presented in an extended field regime (a) and for low magnetic
fields (b). In these measurements the magnetic field was parallel
to the sample in all cases.}
\label{b3}%
\end{figure}

\begin{figure}[tbp] \centering%
\includegraphics[angle=0,width=7cm]{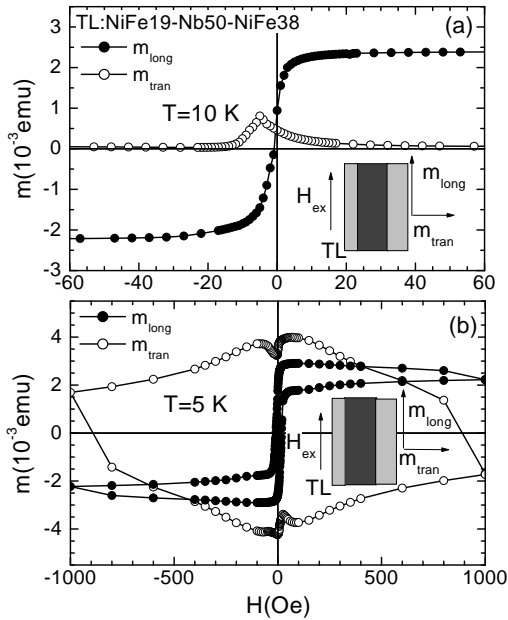}
\caption {Magnetization data obtained in a
NiFe(19)-Nb(50)-NiFe(38) TL (a) at T$=10$ K$>$T$_c^{SC}$ and (b)
at T$=5$ K$<$T$_c^{SC}$ for both the longitudinal (solid circles)
and transverse (open circles) components. The inset in each panel
presents schematically in side view the configuration of the
external field and of each magnetic component of the TL. In these
measurements the external field was parallel to the TL.}
\label{b5}%
\end{figure}

In order to investigate the underlying mechanism responsible for
the magnetoresistance effect we performed detailed magnetization
loop measurements both well below and above the superconducting
transition of Nb. We found out that the m(H) loop of the
NiFe(19)-Nb(50)-NiFe(38) TL obtained at T$=8$ K$>$T$_c^{SC}$ is
almost totally reversible and exhibits nearly zero remanent
magnetization. Detailed results are presented in Figs.\ref{b3}(a)
and \ref{b3}(b) in an extended and in the low-field regime,
respectively. This fact was surprising since we had previously
observed that the respective loops of single NiFe layers exhibit
clear irreversibility and coercivity in the range $10-20$ Oe.
Thus, we investigated this behavior further. Except for the m(H)
loop for the TL (denoted by triangles) we also present data for
two different BLs: a NiFe(19)-Nb(50) (solid circles) and a
Nb(50)/NiFe(38) (open circles). Notice that these two BLs are the
main building blocks of the complete TL. We prepared the two BLs
in order to examine if it was possible to reproduce the m(H) loop
of the TL by decomposing it into its two main parts. This would
allow us to examine whether the two outer NiFe layers of the
complete TL {\it do interact} or the total m(H) loop is a simple
superposition of the two basic m(H) loops. We stress that we
preferred to examine the two NiFe(19)-Nb(50) and Nb(50)-NiFe(38)
BLs and not the two single NiFe(19) and NiFe(38) layers since we
wanted to take into account any possible parameter that could
influence the magnetic behavior of the NiFe layers. For instance,
strain could be induced in each NiFe layer by the adjacent Nb
layer due to the mismatch of their lattices so that the magnetic
m(H) loop of each FM layer could be altered even slightly. The
obtained results are revealing. We clearly see that by adding
appropriately the m(H) loops of the two BLs we may reproduce the
loop of the complete TL only in the high-field regime. The zeroing
of both the coercive field and of remanent magnetization that is
observed in the TL can't be reproduced. Thus, we may safely
conclude that in the low field regime the two NiFe layers
participating the TL strongly interact through the Nb interlayer
so that the magnetic behavior of the complete TL resembles the one
of a soft FM both above and below T$_c^{SC}$ (see
Figs.\ref{b3}(a)-\ref{b3}(b) and Fig.\ref{b5}(b), respectively).
In contrast, for high magnetic fields the interaction between each
NiFe layer with the external magnetic field dominates so that they
practically become uncoupled.

A plausible mechanism that is compatible with the almost zeroing
of the remanent magnetization that is observed only in the TLs is
the antiparallel alignment of the two outer NiFe layers'
magnetizations. However, owing to the specific conditions that
were used during the deposition it is unnatural to assume that our
samples exhibit a {\it coherent} magnetization reversal with
antiparallel relative configuration in the low field regime as it
was observed in Refs.\onlinecite{Pena05,Rusanov06}. In addition,
the Nb interlayer is {\it thick} so that it can't probably
maintain an exchange coupling of the outer NiFe layers. Thus, a
complete antiferromagnetic exchange coupling of the outer NiFe
layers as this occurs in usual giant magnetoresistance spin valves
of {\it thin} normal metal interlayer
\cite{Dieny94,Baibich88,Parkin90} can't be achieved in our case.
Accordingly, other mechanisms should be investigated. Since the
data presented in Fig.\ref{b3} refer to the longitudinal component
of the TL's magnetization (in respect to the external magnetic
field) they don't provide any information on its transverse
component. This information is quite important since the zeroing
of the TL's remanent (longitudinal) magnetization could not be
exclusively related to either an antiparallel alignment of the
outer NiFe layers but could be motivated by the fact that the
magnetization rotates out-of-plane so that it becomes transverse
and can't be detected by the longitudinal set of pick-up coils in
our SQUID. Thus, we performed additional measurements focused on
the transverse magnetic component.

Representative results obtained above and well below T$_c^{SC}$
are shown in Figs.\ref{b5}(a)-\ref{b5}(b) for a TL. Presented are
both the longitudinal (solid circles) and transverse (open
circles) components. We stress that magnetization data for the
superconducting regime, and most importantly for the transverse
magnetic component, are presented for the first
time.\cite{Gu02,Moraru06,Pena05,Rusanov06,StamopoulosPRB06new} The
insets present the configuration of the external field and of each
magnetic component. Let us first discuss panel (a), that is the
normal state data obtained at T$=10$ K$>$T$_c^{SC}$ while the
applied field is lowered from positive saturation. We clearly see
that as the external field passes through zero the transverse
magnetic component of the TL attains significant values of the
order $40\%$ of the saturated longitudinal one. This clearly
proves that in our samples a significant part of the TL's
magnetization rotates out-of-plane near zero magnetic field.
Proceeding with the data of panel (b), that were obtained in the
superconducting state at T$=5$ K$<$T$_c^{SC}$ we stress that these
data reveal a surprising feature: the out-of-plane rotation of the
TL's magnetization is not observed only in the normal state but
also well inside the superconducting state of Nb. We see that even
at T$=5$ K the longitudinal component resembles the loop of a FM
as if the SC was absent (the only fingerprint of its presence
comes from the comparatively small irreversibility that shows up).
More importantly, we see that it is the transverse component of
the TL that obtains the model loop expected for a SC. {\it These
results clearly prove that the SC behaves diamagnetically not in
respect to the parallel external field but in respect to a new
transverse field that emerges owing to the magnetic coupling of
the outer NiFe layers.}

\subsection{Comparison of low field data}

\begin{figure}[tbp] \centering%
\includegraphics[angle=0,width=7cm]{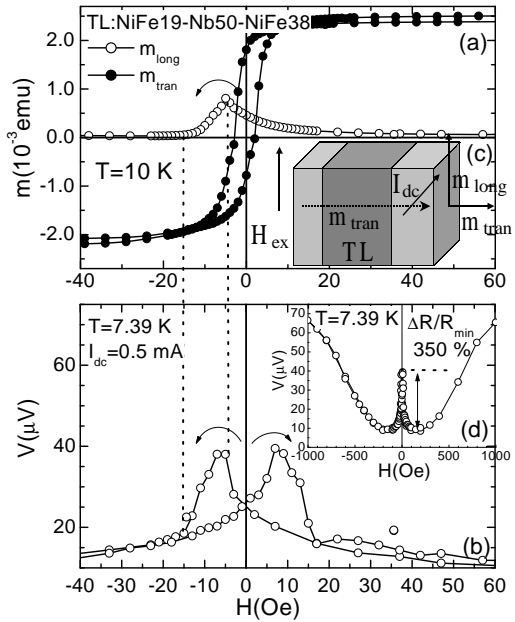}
\caption {(a) Both branches of the longitudinal (solid circles)
and decreasing branch of the transverse (open circles) magnetic
components obtained at T$=10$ K$>$T$_c^{SC}$ and (b) both branches
of the magnetoresistance curve obtained at T$=7.39$ K$<$T$_c^{SC}$
for a NiFe(19)-Nb(50)-NiFe(38) TL. Inset (c) presents
schematically the configuration of both the magnetization and the
transport measurements, while inset (d) shows the
magnetoresistance curve in an extended field range. In these
measurements the external field was parallel to the TL.}
\label{b6}%
\end{figure}

The results that were presented above indicate that probably the
out-of-plane rotation of the TL's magnetization is related to the
observed magnetoresistance peaks. To investigate this possibility
thoroughly in Figs.\ref{b6}(a) and \ref{b6}(b) we compare detailed
measurements of the longitudinal (solid circles) and the
transverse (open circles) magnetic components (obtained at T$=10$
K$>$T$_c^{SC}$) and of the magnetoresistance (obtained at T$=7.39$
K$<$T$_c^{SC}$), respectively. These data are focused at low
magnetic fields. Inset (c) shows schematically the configuration
for both the magnetization and transport measurements, while inset
(d) shows the magnetoresistance curve in an extended field range.
We clearly see that {\it the maximum of the TL's transverse
component coincides with the magnetoresistance peak}. In addition,
the zeroing of the transverse magnetization is accompanied by an
abrupt change in the slope of the V(H) curve. {\it This
experimental fact indicates that in the TLs the out-of-plane
rotation of the magnetization is probably responsible for the
observed magnetoresistance peaks.}

\begin{figure}[tbp] \centering%
\includegraphics[angle=0,width=7cm]{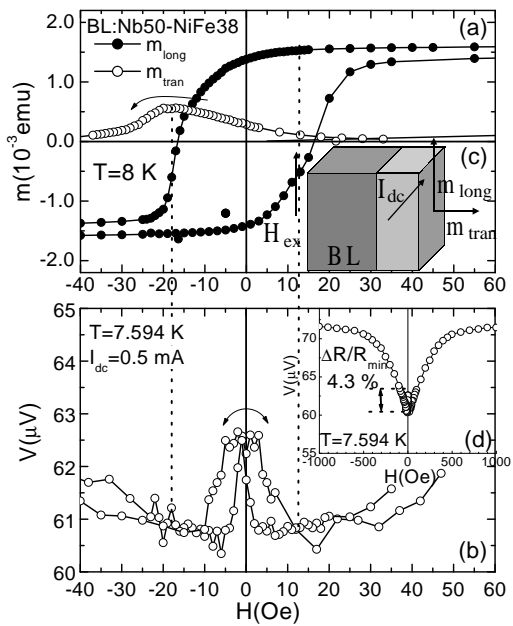}
\caption {(a) Both branches of the longitudinal (solid circles)
and decreasing branch of the transverse (open circles) magnetic
components obtained at T$=10$ K$>$T$_c^{SC}$ and (b) both branches
of the magnetoresistance curve obtained at T$=7.594$
K$<$T$_c^{SC}$ for a Nb(50)-NiFe(38) BL. Inset (c) presents
schematically the configuration of both the magnetization and the
transport measurements, while inset (d) shows the
magnetoresistance curve in an extended field range. In these
measurements the external field was parallel to the BL.}
\label{b7}%
\end{figure}%

To investigate if this holds also for the case of BLs we performed
the respective measurements in the low field regime.
Representative results of the longitudinal (solid circles) and
transverse (open circles) magnetic components (obtained at T$=8$
K$>$T$_c^{SC}$) and of the magnetoresistance (obtained at
T$=7.594$ K$<$T$_c^{SC}$) are shown in Figs.\ref{b7}(a) and
\ref{b7}(b), respectively. Inset (c) shows schematically the
configuration for both the magnetization and transport
measurements, while inset (d) shows the magnetoresistance curve in
an extended field range. The direct comparison reveals that the
maximum of the BL's transverse component follows the zeroing of
the longitudinal one. More importantly, we see that {\it in the
BLs the minor magnetoresistance peak is clearly not related to the
out-of-plane rotation of the magnetization since it is placed at
almost zero magnetic field}.

\section{Discussion}

Summarizing the data presented so far we conclude that: (i) The
magnetoresistance effect observed in the TLs is one order of
magnitude stronger than the one of the BLs (see
Figs.\ref{b2}(a)-\ref{b2}(b)). (ii) When the NiFe layers are
brought together through a Nb interlayer they strongly interact so
that almost zeroing of both the longitudinal remanent
magnetization and coercivity is observed (see Fig.\ref{b3}). In
contrast, the height of the transverse magnetic component is not
seriously affected except for it becomes more sharp in the TLs
($\Delta$ H$\simeq 30$ Oe) when compared to the BLs ($\Delta$
H$\simeq 50$ Oe) as may be easily seen in Figs.\ref{b6}(a) and
\ref{b7}(a), respectively. This fact indicates that in the TLs the
out-of-plane rotation is forced by the interaction of the outer
NiFe layers. (iii) The SC exhibits transverse magnetization
component since it interacts rigidly with the outer FM layers (see
Fig.\ref{b5}(b)). (iv) Finally, although the out-of-plane rotation
of the magnetization is surely related to the strong
magnetoresistance effect that is observed in the TLs (the relative
peaks of the m$_{tran}$(H) and V(H) curves presented in
Figs.\ref{b6}(a) and \ref{b6}(b), respectively clearly coincide),
for the BLs it has no or minor influence (the relative peaks of
the m$_{tran}$(H) and V(H) curves presented in Figs.\ref{b7}(a)
and \ref{b7}(b), respectively are clearly not related). Based on
these experimental facts below we discuss the most possible
mechanism for their interpretation. We discuss first the TLs where
the effect is peculiar.


As stated above, reason (iv) indicates that in the TLs the
magnetization's out-of-plane rotation surely contributes to the
observed effect. However, we believe that in the TLs an extra
ingredient should also exist that promotes the magnetoresistance
effect so strongly since in the BLs the magnetization's
out-of-plane rotation also occurs but the accompanied
magnetoresistance effect is only minor. This extra ingredient
comes from reasons (ii) and (iii): in the TLs it is an {\it
interaction} mechanism between the outer NiFe layers that is
getting involved. Since the Nb interlayer is rather thick we
believe that this interaction refers rather to a stray-fields
induced magnetostatic coupling of the outer NiFe layers than an
exchange one. On the other hand, although in the BLs this
mechanism could also be involved, it surely has a mild
contribution since the single NiFe layer doesn't have the
opportunity to get coupled with a second one so that the Nb
interlayer is not pierced effectively by the transverse stray
fields. Here we present a complete discussion on the possibility
of such magnetostatic coupling.

It is well known that in relevant TLs of normal or insulating
interlayer the interaction of the outer FM layers through stray
fields that occur at domain walls may lead to significant
magnetostatic coupling.\cite{Parkin98,Parkin00} This behavior is
expected to be pronounced when the FM layers have a multi-domain
magnetic state, that is in the low magnetic field regime, near
coercivity. S. Parkin and colleagues have shown
\cite{Parkin98,Parkin00} that such stray-fields coupling that
occurs at domain walls plays a unique role in FM-IN-FM and
FM-NM-FM TLs (IN and NM stand for insulator and non-magnetic
metal, respectively) since it vitiates potently the distinct
magnetic character of even quite different outer FM layers.
Accordingly, in our case the partial out-of-plane rotation of the
TL's magnetization that we observe in the normal state (see
Fig.\ref{b5}(a)), and most importantly {\it the unique
out-of-plane rotation of the SC's magnetization that we observe in
the superconducting state} (see Fig.\ref{b5}(b)) are probably
motivated by such a stray-fields magnetostatic coupling of the
outer NiFe layers: in consistency with the mechanism proposed in
Refs.\onlinecite{Parkin98,Parkin00} the out-of-plane rotation that
we observe in our data occurs progressively in the low field
regime where as the zeroing of the longitudinal component implies
many domains get formed so that this mechanism could be activated.

\begin{figure}[tbp] \centering%
\includegraphics[angle=0,width=6.5cm]{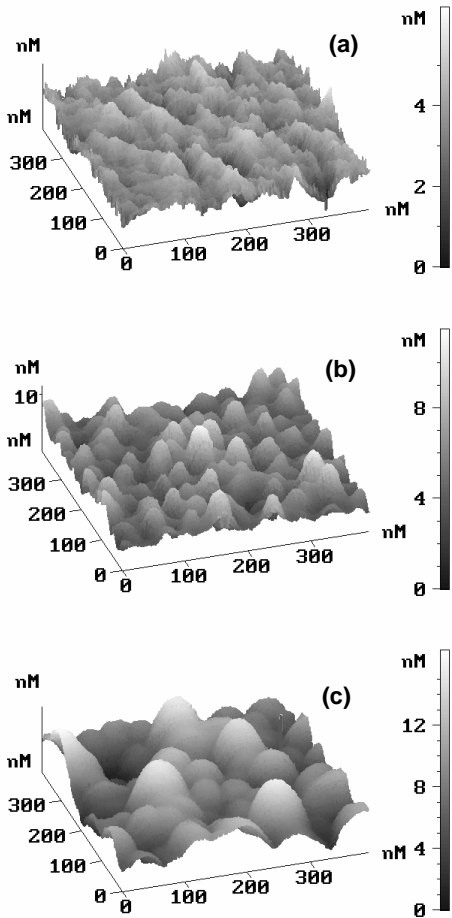}
\caption {Representative AFM images obtained in (a) a NiFe(19)
layer, (b) a NiFe(19)-Nb(50) BL, and (c) a
NiFe(19)-Nb(50)-NiFe(38) TL. In all cases the presented areas are
$0.4\times0.4$ $\mu$m$^2$. Notice the different scales in the
respective vertical black-and-white bars among panels (a)-(c).}
\label{AFM}%
\end{figure}%

\begin{table}[tbp] \centering%
\caption{Peak-to-peak, mean and root-mean-square roughness values
for samples NiFe(19) SL19, NiFe(19)-Nb(50) BL and
NiFe(19)-Nb(50)-NiFe(38) TL.}
\begin{ruledtabular}
\begin{tabular}{ccccc}
$  $       &peak-to-peak (nm)       &mean (nm)      &root-mean-square (nm)\\
$SL19$     &6.3             &3.2            &0.7\\
$BL$       &11.3            &4.9            &1.4\\
$TL$       &16.7            &8.1            &2.8\\

\end{tabular}
\end{ruledtabular}
\label{table}
\end{table}

The occurrence of such stray-fields coupling between the outer
NiFe layers could also be promoted when significant roughness
exists at the
interfaces.\cite{Parkin98,Parkin00,Parkin00b,Egelhoff06} In order
to investigate this possibility we performed a thorough AFM study
of our samples. Representative results are shown in
Figs.\ref{AFM}(a)-\ref{AFM}(c). Panel (a) refers to a NiFe(19)
single layer (SL), panel (b) to the top surface of a
NiFe(19)-Nb(50) BL, and panel (c) to the top surface of a
NiFe(19)-Nb(50)-NiFe(38) TL. In all cases the presented areas are
$0.4\times0.4$ $\mu$m$^2$. We studied all these structures because
we wanted to examine if the surface roughness presents an
evolution from layer to layer. Notice that the scale in the
respective vertical black-and-white bars is different among panels
(a)-(c). The obtained values for each sample are displayed in
Table \ref{table}. These data reveal that the roughness of the
interfaces exhibits an evolution as a new layer is progressively
added to the previous one(s). Thus, the magnetostatic coupling of
the outer NiFe layers could be enhanced through local "guidance"
of the stray fields by the roughness of the interfaces.
\cite{Parkin98,Parkin00,Parkin00b,Egelhoff06}

Ultimately, the magnetostatic coupling discussed above could
motivate the magnetoresistance peaks that we observe as following:
the stray fields emerge naturally near coercivity owing to the
attained multidomain magnetic state so that they interconnect
vis-a-vis domains that are hosted in the outer NiFe layers. Since
these stray fields evolve in the whole surface of the NiFe layers
they penetrate completely the Nb interlayer (see the schematic
inset (c) in Fig.\ref{b6}). Depending on the specific
characteristics of the employed FM layers (exchange field, size of
magnetic domains, width and kind -Neel or Bloch- of domain walls,
etc) and of the SC interlayer (lower and upper critical field
values, bulk pinning force, etc) the FMs' stray fields will
primarily exceed the SC's lower critical field and secondarily, in
case that they are intense, even the SC's upper critical field
could {\it locally} be exceeded, at least extremely close to its
T$_c^{SC}$ where it attains low values. These two different cases
are discussed in the two following paragraphs.

{\it Case (a):} In case where only the SC's lower critical field
is exceeded by the FMs' stray fields dissipation should set in due
to the following reason: since the magnetostatic coupling of the
outer NiFe layers leads to a transverse magnetization field that
penetrates entirely the Nb interlayer we may assume that vortices
enter the Nb interlayer in the form of "chains" that are mainly
positioned above domain walls. Thus, the SC interlayer is in the
mixed state. The current applied during our transport measurements
is normal to such "chains" of vortices so that it exerts a Lorentz
force on them.\cite{Tinkham} This leads to movement of vortices
which is well known that results in dissipation.\cite{Tinkham}
Eventually, as we lower the temperature the magnetoresistance
peaks become progressively smaller since bulk pinning sets in and
vortices are no longer free to move, or the increased lower
critical field exceeds the stray fields so that a dissipationless
Meissner state is ultimately recovered.\cite{Tinkham} This
explanation resembles the one that was presented by V.V. Ryazanov
et al. in Ref.\onlinecite{Ryazanov03} for
Nb-Cu$_{0.43}$Ni$_{0.57}$ BLs.

{\it Case (b):} In this case we may assume that since the
pronounced magnetoresistance effect is observed extremely close to
T$_c^{SC}$ where except for the SC's lower critical field, its
upper critical field is also very low, the FMs' stray fields could
even {\it locally} overstep it. Thus, {\it localized} normal areas
of the SC that lie above FM areas where intense stray fields
emerge should contribute extra dissipation. Eventually, in the
scenario discussed here the disappearance of the magnetoresistance
peaks is owing to the fact that as we progressively lower the
temperature the SC's upper critical field exceeds the stray fields
that interconnects the outer FMs so that bulk superconductivity is
completely restored throughout the whole SC interlayer. Of course,
since the lower critical field is always smaller than the upper
critical one we expect that the mechanism described in case (a)
(case(b)) should have a major (minor) contribution to the observed
magnetoresistance effect.

Here let us compare our results with current experiments that are
very relevant to our work. Regarding the TLs, we believe that the
specific characteristics of our films owing to their preparation
process (soft magnetic behavior, no uniaxial magnetic anisotropy,
noticeable interface roughness that evolves from layer to layer)
explain the differences in both the magnitude of the
magnetoresistance and the proposed mechanisms between the TLs
studied in Refs.\onlinecite{Pena05,Rusanov06} and in our work. A.
Yu. Rusanov et al. reported \cite{Rusanov06} on the
magnetoresistance of similar NiFe-Nb-NiFe TLs and attributed the
magnetoresistance peaks that they observed to the antiparallel
alignment of the outer NiFe layers. Their samples were deposited
while the substrates were directed along the sputtering chamber's
residual fields so that they exhibit strong uniaxial anisotropy
with the easy axis of magnetization placed along their long
dimension. Thus, in their case as the magnetic field passes
through zero the magnetization of each NiFe layer participating a
TL reverses {\it coherently} and probably in-plane without
exhibiting significant out-of-plane component (although data on
the transverse magnetic component are not presented in
Ref.\onlinecite{Rusanov06}). According to the same arguments the
significant difference between the magnetoresistance values
observed in Ref.\onlinecite{Rusanov06} and in our work is also
explained. In Ref.\onlinecite{Rusanov06} it is reported that by
using the definition $(R_{max}-R_{min})/R_{nor}\times100\%$ the
maximum magnetoresistance value amounts to $5\%-10\%$, while in
our case the respective value is $45\%$ (see Fig.\ref{b2}(a)) for
the TL presented in this work (in other samples we even observed
values exceeding slightly $50\%$). Also, V. Pe\~{n}a et al.
\cite{Pena05} have reported on a magnetoresistance effect observed
in
La$_{0.7}$Ca$_{0.3}$MnO$_3$-YBa$_2$Cu$_3$O$_7$-La$_{0.7}$Ca$_{0.3}$MnO$_3$
TLs. By using the definition
$(R_{max}-R_{min})/R_{min}\times100\%$ the authors noted a
magnetoresistance increase of $1000\%$ in the superconducting
state. Thus, quantitatively the effect that we observe (see
Fig.\ref{b2}(a)) is identical to the one reported in
Ref.\onlinecite{Pena05}. This is quite surprising since both
constituents La$_{0.7}$Ca$_{0.3}$MnO$_3$ and YBa$_2$Cu$_3$O$_7$
have very different properties from the materials used in our TLs.
While La$_{0.7}$Ca$_{0.3}$MnO$_3$ is highly spin polarized (almost
$100\%$) NiFe has a comparatively low spin polarization ($45\%$).
Also, YBa$_2$Cu$_3$O$_7$ is a representative high-T$_c$ material,
while Nb is a well studied low-T$_c$ one. Consequently, we may
assume that the observation of a similar (if not identical)
magnetoresistance effect in both
Ni$_{80}$Fe$_{20}$-Nb-Ni$_{80}$Fe$_{20}$
(Ref.\onlinecite{Rusanov06} and this work) and
La$_{0.7}$Ca$_{0.3}$MnO$_3$-YBa$_2$Cu$_3$O$_7$-La$_{0.7}$Ca$_{0.3}$MnO$_3$
(Ref.\onlinecite{Pena05}) TLs could imply a generic underlying
mechanism that doesn't depend on specific characteristics as the
spin polarization of the FM and the pairing mechanism of the SC.
Examining the influence of the transverse magnetic component in
all relevant TLs could shed more light on this issue.
Nevertheless, the results obtained in Refs.
\onlinecite{Pena05,Rusanov06} and in our work are attractive for
the design of low-field sensor devices.

Finally, regarding the BLs a stray-fields mechanism analogous to
the one existing in the TLs is surely getting involved but with
much smaller intensity since the single NiFe layer doesn't have
the opportunity to get coupled magnetostatically with an adjacent
one. Thus, now the stray fields don't penetrate completely the Nb
interlayer so that they have only a minor influence on it.

\section{Summary and Conclusions}

In summary, we demonstrated that TLs comprised of low spin
polarized NiFe and low-T$_c$ Nb exhibit a pronounced
magnetoresistance effect: the magnetoresistance increases two
orders of magnitude when the superconducting state is reached,
from the normal-state value $\Delta R/R_{nor}\times100\%=0.6\%$ it
goes up to $\Delta R/R_{nor}\times100\%=45\%$ ($\Delta
R/R_{min}\times100\%=1000\%$) for temperatures below T$_c^{SC}$.
In contrast, the BLs exhibit a minor effect: from $\Delta
R/R_{nor}\times100\%=3\%$ in the normal state the
magnetoresistance increases only to $\Delta
R/R_{nor}\times100\%=8\%$ ($\Delta R/R_{min}\times100\%=70\%$).

Magnetization data referring to the evolution of the transverse
component from the normal to the superconducting state are
presented for the first time. Regarding the normal state these
magnetization data revealed that in the same field range where the
magnetoresistance peaks are observed the TL's magnetization
partially rotates out-of-plane. Surprisingly, in the
superconducting regime the SC's magnetization follows rigidly the
outer FM layers and also exhibits a transverse component. We
attribute these facts to the magnetostatic coupling of the outer
NiFe layers owing to the stray fields that, near coercivity,
evolve naturally in their whole surface. As a result the
"magnetically pierced" SC interlayer behaves diamagnetically not
in respect to the longitudinal external magnetic field but in
respect to these stray fields that interconnect the outer NiFe
layers.

Thus, the strong magnetoresistance effect that is observed in the
TLs around coercivity is owing to the stray fields that as they
penetrate entirely the Nb interlayer they exceed either only its
lower or even its upper critical field. This detail can't be
elucidated by our measurements. However, in the general case this
should depend on the specific characteristics of the employed FM
and SC constituents.

Referring to the BLs although out-of-plane rotation of the single
NiFe layer is observed in the magnetization measurements, the
accompanying stray fields have only a modest influence on
magnetoresistance since magnetostatic coupling is not realized
owing to the absence of a second NiFe layer.

{\it Note added:} During the review process we become aware of the
article [R. Steiner and P. Ziemann, Phys. Rev. B {\bf 74}, 094504
(2006)] which deals with the same subjects. R. Steiner and P.
Ziemann study hybrid TLs and BLs consisting of Co, Fe and Nb by
performing transport and {\it longitudinal} magnetization
measurements. However, the measurements of the longitudinal
magnetic component were limited in the normal state. In addition,
the authors also employ micromagnetic simulations to convincingly
show that in their samples the stray fields are related to the
observed dissipation peaks. In our work we employ magnetization
measurements of both the {\it longitudinal} and {\it transverse}
magnetic components in the TLs and their building-blocks BLs to
reveal the contribution of this mechanism. Our magnetization data
extend in the superconducting state uncovering important
information on the coupling between the outer FM layers and the SC
interlayer. Obtained by different means the basic explanation
proposed in [R. Steiner and P. Ziemann, Phys. Rev. B {\bf 74},
094504 (2006)] and in our work is the same.

\pagebreak

\end{document}